\documentclass{optica-article}
\journal{oe}

\usepackage{lineno}

\usepackage{color}

\usepackage{subcaption}

\newdimen\figrasterwd
\figrasterwd\textwidth

\usepackage{mathrsfs}
\usepackage{cite}
\usepackage{epsf}
\usepackage{theorem}
\usepackage{times} 
\usepackage{graphics} 
\usepackage{epsfig} 
\usepackage{siunitx}

\usepackage{soul}

\newcommand{\DMB}{DSCM}

\newcommand{\LSTM}{LSTM}
\newcommand{\DMBCombinedFC}{Common-Core}
\newcommand{\DMBCombinedFCabbr}{CC}
\newcommand{\DMBCombinedPerTone}{Separate-Core-per-Band}
\newcommand{\DMBCombinedPerToneabbr}{SC}
\newcommand{\DMBvTwo}{Modular-I}  
\newcommand{\DMBvTwoabbr}{M1}

\newcommand{\DMBModular}{Modular-II}
\newcommand{\DMBModularabbr}{M2}

\begin{document}

%

\title{Neural Network Architectures for Optical Channel Nonlinear Compensation in Digital Subcarrier Multiplexing Systems}

\author{Ali Bakhshali\authormark{*}, Hossein Najafi, Behnam Behinaein Hamgini, and Zhuhong Zhang} 

\address{Ottawa Optical Competency Center, Huawei Technologies Canada, 303 Terry Fox Dr, Kanata K2K 3J1}

\email{\authormark{*}ali.bakhshali@huawei.com}

%
%

\begin{abstract*}
In this work, we propose to use various artificial neural network (ANN) structures for modeling and compensation of intra- and inter-subcarrier fiber nonlinear interference in digital subcarrier multiplexing (DSCM) optical transmission systems. 
%
We perform nonlinear channel equalization by employing different ANN cores including convolutional neural networks (CNN) and long short-term memory (LSTM) layers.  
We start to compensate the fiber nonlinearity distortion in \DMB~systems by a 
fully-connected network across all subcarriers. 
In subsequent steps, and borrowing from fiber nonlinearity analysis, we gradually upgrade the designs towards modular structures with better performance-complexity advantages. 
Our study shows that putting proper macro structures in design of 
ANN nonlinear equalizers in \DMB~systems can be crucial for practical solutions in future generations of coherent optical transceivers.
\end{abstract*}


\section{Introduction}
For high-speed long-haul fiber-optic transmission, the nonlinear interference from Kerr effect is a major bottleneck that limits the achievable transmission rates. 
This interference can be equalized by approximating and inversing 
the nonlinear Schrodinger equation through digital back-propagation (DBP) 
\cite{DBP1, DBP2, DBP3} or perturbation-based nonlinear compensation (PNLC) \cite{pert1, pnlc3}. 
However, large computational complexities and also the need for accurate data from the propagation link have limited their application in real-time processing with agile and flexible requirements.
%

Alternatively, a variety of ANN solutions have been recently proposed for fiber nonlinearity compensation application. 
The primitive works tried to squeeze additional performance by feeding triplets inspired from perturbation analysis of fiber nonlinearity to a feed-forward neural network\cite{NLC-MLP1}. Later works draw inspiration from DBP and aimed to incorporate deep convolutional neural networks (CNNs) for this tasks \cite{LDBP1,CNN1}. The use of advance recurrent neural networks (RNNs), such as long short-term memory (LSTM) modules, which are more suitable for the equalization of time-series processes has also picked up a great interest \cite{deligiannidis2020compensation,perf-comp1}. 
In fact, the pattern and medium dependent characteristics of nonlinear propagation make it a suitable problem to be tackled by variety of solutions from artificial neural network domain.
In general, an ANN-based nonlinear equalizer is more flexible compared to the conventional methods in the sense that it can be better updated for different transmission scenarios without the need for accurate feedback of the channel parameters. Also, ANN nonlinear equalizers can be extended to include the functionalities of traditional DSP modules to create a more general equalizer.
Furthermore, the ANN design where the compensation process is learned through data can potentially lead to a large reduction in computational complexity
\cite{QNN1}.

In this work, we consider an application of ANN in coherent optical communications. We focus on advanced ANN structures with the ability to generate appropriate features without any reliance on an external module for feature generation. We particularly study digital subcarrier multiplexing (\DMB)~systems since their design flexibility makes it a promising solution for the coherent optical modems. Simplifying the DSP with lower speed processing per-subcarrier, flexible channel-matched transmission, robust clock recovery, and the easy transition to a point-to-multi-point (P2MP) architecture are some of the advantages of \DMB~systems.

Here, we develop \emph{macro} ANN structures, inspired by the fiber nonlinearity distortion mechanism that governs the nonlinear interaction across different subcarriers that has shown to be more efficient in terms of  inference complexity, model representation, and the ability to be trained.
We propose various ANN structures for modeling and compensation of intra- and inter-subcarrier fiber nonlinearities in \DMB~systems, and explore 
scalability and performance versus complexity tradeoffs of the presented solutions. 
Different models are designed in terms of how received symbols across digital subcarriers are employed for training ANN cores for intra-subcarrier self-phase modulation (iSPM) and inter-subcarrier cross-phase modulation (iXPM) nonlinear impairments. 
Starting with a fully-connected network across all subcarriers, we move toward 
upgrading the design with modular ANN cores and sequential training stages.
In other words, we start with black-box ANN models and then propose more efficient and flexible modular designs inspired by nonlinear perturbation analysis.
All models in here are universal from ANN-core choice perspective.
Specifically, we choose the building block for all the proposed structures in this work to be an ANN core with combinations of CNN and LSTM layers.
One important aspect in this work is to generalize the neural network designs such that a block of data is generated in the equalization since parallelization is an essential feature of the coherent modems. 
We explore parallelization of these designs and impact of block-processing on performance versus complexity tradeoffs for these models. 
We show that one can get orders of magnitude reduction in computational complexity by moving towards block equalization in this fashion. 

The remainder of this paper is organized as follows: In 
Section~\ref{NLC_pre_sec}, the base of nonlinear compensation for fiber channel is briefly discussed. 
In Section~\ref{MP_NN_core}, the multi-purpose ANN-core structure as the building block of proposed models is explained.
The details of various ANN structures for NLC in \DMB~are presented Section~\ref{NN_struct_sec} while Section~\ref{sys_dmb} is devoted to the numerical setup and results comparison.
Next, impact of dispersion map on the nonlinear equalizer designs is discussed in Section~\ref{Rx-side CDC}. 
Finally, we conclude the paper in Section~\ref{conclusion_sec}.


\section{Nonlinear Compensation for Optical Fiber Channel}\label{NLC_pre_sec}
The dual-polarization evolution of optical field over a fiber link can be explained by the Manakov equation \cite{Manakov} where the linear and nonlinear propagation impacts are described as follows:
\begin{align}
\frac{\partial{u_{x/y}}}{\partial{z}} + \frac{\alpha}{2}u_{x/y} + j\frac{\beta}{2}\frac{\partial^2u_{x/y}}{\partial{t^2}} = j\frac{8}{9}\gamma\Bigl[|u_x|^2+|u_y|^2\Bigr]u_{x/y},
\label{mana_eq}
\end{align}
where $u_{x/y} = u_{x/y}(t,z)$ represents the optical filed of polarization $x$ and $y$, respectively, $\alpha$ is the attenuation coefficient, $\beta$ is the group velocity dispersion (GVD), and $\gamma$ is the nonlinear coefficient.
The nonlinear interference can be equalized by approximating and inversing the 
above equation through DBP \cite{DBP1, DBP2, DBP3} where the fiber is modeled 
as a series of linear and nonlinear sections through first-order approximation 
of Manakov equation.  
On the other hand, by employing the perturbation analysis \cite{pert1}, one can represent the optical field as the solution of linear propagation plus a perturbation term from the nonlinear impact in symbol domain and in one step for the accumulated nonlinearities. It is shown that the first-order perturbation term can be modeled by the weighted sum of triplets of transmitted symbols plus a constant phase rotation\cite{pnlc3, pnlc4}.

\begin{figure}[t]  
	\centering	
	\includegraphics[trim={2cm 2cm 1.8cm 1.5cm},clip, width=0.9\textwidth]{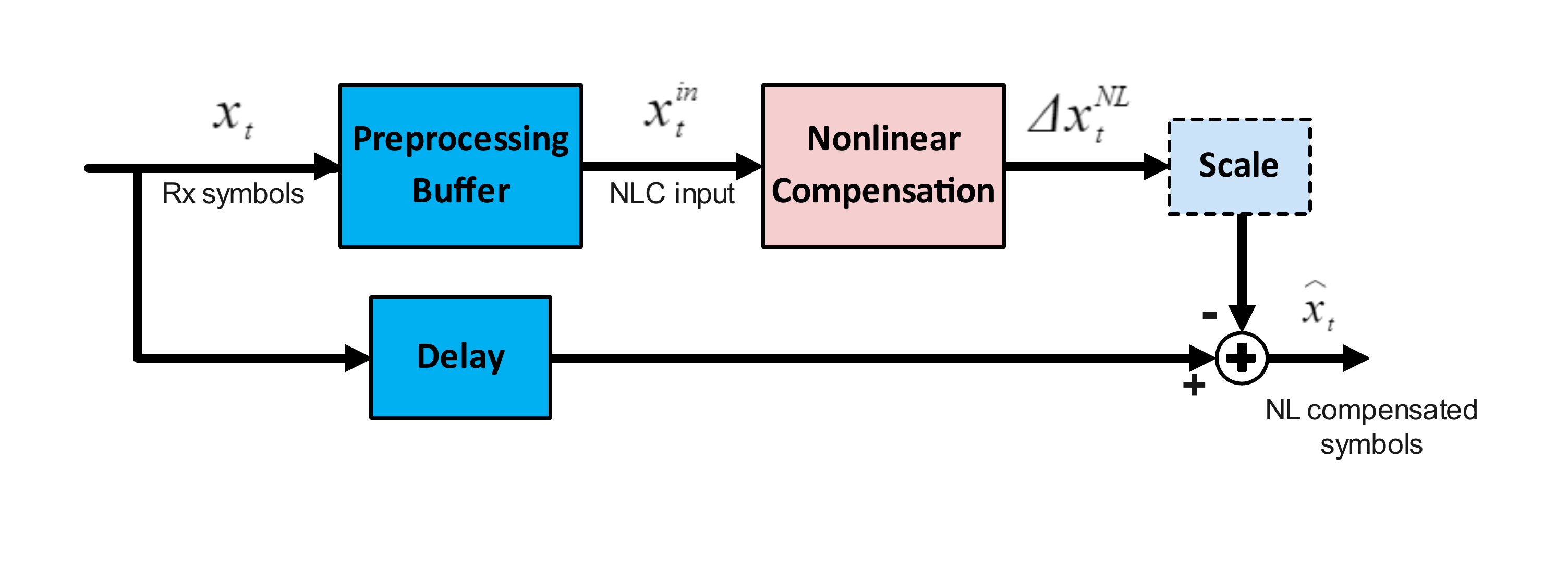}
	\caption{Block diagram for lumped perturbation-based nonlinear compensation.}
	\label{diag:ANN}
\end{figure}

A wide variety of machine learning solutions for fiber nonlinearity compensation in optical communications has been proposed in literature (\cite{NLC-MLP1, LDBP1, CNN1, perf-comp1} among others). These solutions are generally benchmarked against conventional solutions such as DBP and PNLC.
Considering the lumped nonlinear compensation methods, a block diagram for the 
equalization module is presented in Fig.~\ref{diag:ANN} where the pre-processing buffer generates appropriate inputs for a given method. 
Specifically, it includes a module that calculates appropriate PNLC triplets  
for regular perturbation-based method or an artificial neural network nonlinear compensation (ANN-NLC) approach that operates on externally generated triplet features \cite{NLC-MLP1}. In ANN-NLC solutions that directly operate on Rx-DSP outputs \cite{CNN1,	perf-comp1,	NN_Optic,co-lstm}, this pre-processing buffer is tasked to provide an extended block of soft symbols needed to efficiently equalize the nonlinear interference.

Considering the first-order perturbation as the dominant nonlinear term, an 
appropriate scaling can be employed to adapt the nonlinear error estimates 
in case the training and inference stages are performed at different optical launch powers: 
\begin{equation}
\alpha = 10^{\bigl(P_\text{inference}(\text{dB}) - 
P_\text{train}(\text{dB})\bigr)/10}.
\label{ANN_scale}
\end{equation}
where $P_\text{train}$ is the optical launch power of the data used in the training while $P_\text{inference}$ is the respective optical launch power of data in the inference (equalization) stage.

In this work, we consider various lumped ANN structures for modeling and 
compensation of fiber nonlinearity for \DMB~systems. 
%
We provide the evolutionary approach of designing advanced ANN models that do not rely on external features (such as triplets) as input. 
Hence, by using symbols in a delay-line format as the input, the model learns relevant features according to its structures through additional layers.
The proposed ANN-NLC equalizers estimate the nonlinear distortions of each 
subcarrier in one polarization of a \DMB~signal given the relevant information from all digital subcarriers across both polarizations. 
Due to the nature of signal propagation in fiber and symmetries in the medium, it has been shown that the same model can be used to generate nonlinear error estimates for other polarization by simply swapping input signals to their respective counterpart from the first polarization. This alleviates the need to train separate models for X and Y polarization and enables efficient learning of a generalized model.

\section {Multi-Purpose ANN-Core Structure}\label{MP_NN_core}
The presented ANN networks here mainly explore different higher level structures that aim to exploit the interaction between each target digital subcarrier and the neighboring ones in search for more powerful and efficient models. 
Hence, the models are universal from ANN-core choice perspective. 
Specifically, we choose the building block for all the proposed models in this work to be an ANN core comprising a combination of CNN and LSTM cores. 
In particular, we employ LSTM units that has been shown to have promising capability in efficient learning of complicated dynamic nonlinear systems. 
The first layer is a 1-dimensional CNN followed by Leaky ReLU activation function that is tasked with helping in feature generation. CNN features are fed into a \LSTM~module with bi-directional structure to extract time-dependency of the input features. 

To save in computational resources for equalization, one can share the computational overhead corresponding to initialization of each LSTM chain by expanding the input and output sequences and provide estimates for multiple time instances. LSTMs are highly suitable to reduce the processing overhead of a sequential input stream since they aim to capture the most relevant representations of the past observed inputs in form of the hidden states. These hidden state variables are updated as new inputs are processed sequentially. However, the output remains an explicit function of the inputs and hidden state variables at every time instance. 
Consequently, equalization of any extra input only increases the total computations by one extra RNN process step. To leverage this capability, simplify training, and to avoid the challenges from long 
back-propagation in LSTMs, these neural networks are trained with the regular symbol-based processing while the block-processing is employed during the deployment and evaluation stage.
Note that by using the block-processing in equalization path, we bring an 
approximation into the network that we have trained with different initial 
hidden states. 
However, with long enough training block size and filter-tap in LSTM, one can 
show that the change in the states are minimal \cite{co-lstm}. 
This is reflected in the complexity figures as we deploy trained models with different block sizes $N$ in the numerical results.

\begin{figure}[t]
	\centering	
	\includegraphics[trim={1cm 1cm 1cm 1cm},clip, width=0.85\textwidth] 
	{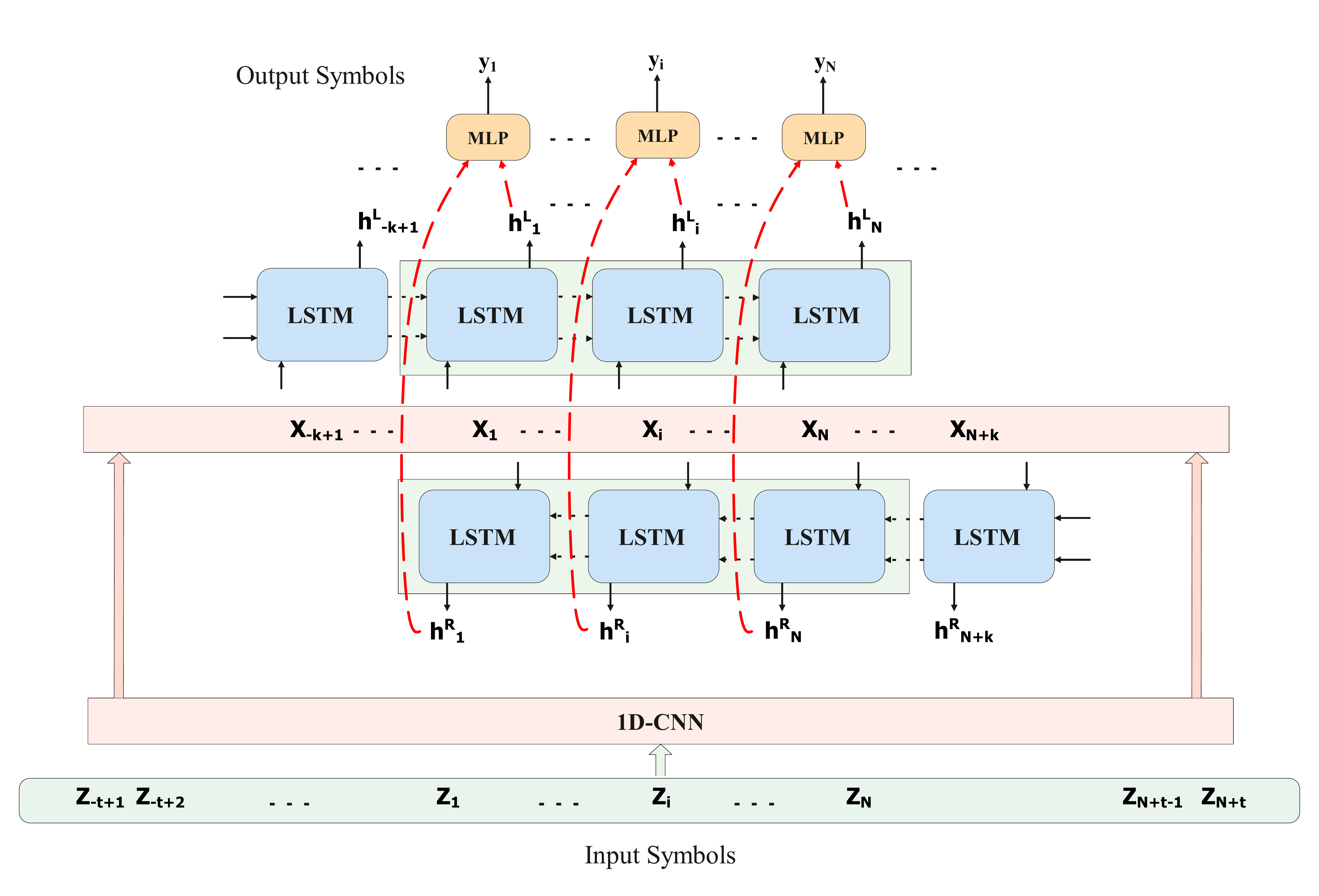}	
	\caption{Multi-purpose ANN-Core structure for the equalization path.}
	\label{diag:vector_lstm_0}
\end{figure}

A block diagram for the proposed ANN equalization core is depicted in Fig.~\ref{diag:vector_lstm_0}. 
The LSTM network has been trained using a fixed sequence of features corresponding to $2k+1$ time instances, where $k$ is the filter-tap size on each side of the target symbol.
In the equalization path, we deploy the same network over input feature sequences corresponding to $2k+N$ time instances to obtain output features associated to the symbols in the middle $N$ time instances. 
In this case, input features corresponding to the first $k+N$ time-instances $i \in\{-k+1,\ldots,N\}$ are sequentially fed into a forward LSTM unit initialized with zero memory, producing output features and evolving the internal memory states.
A similar, backward LSTM unit starts with zero memory and evolves using the CNN features corresponding to the last $k+N$ time instances of the $2k+N$ window $i \in\{1,\ldots, N+k\}$ in the opposite direction. Outputs of forward and backward LSTM modules for middle $N$ time instances are concatenated to form the LSTM block outputs.
Finally, the LSTM block outputs may pass through a linear or a multi-layer 
perceptrons (MLP) stage with Leaky ReLU activation functions (for all but the 
last layer) that ultimately 
provides estimations for real and imaginary part of nonlinear interference per 
output. 
Note that, as we discuss further in section \ref{NN_struct_sec}, the final MLP 
layer can be separated from the ANN core and trained individually in some 
architectures.

In order to get a measure of complexity for the multi-purpose ANN core, we 
consider a CNN-LSTM network with an MLP output layer. Let us consider equalization of $N$ symbols with processing window of $N_{w} = 2t+N$. 
The real multiplication per symbol (RM) for CNN is equal to:
\begin{align}
	CNN_{RM} = \frac{4N_fN_{ke}(N_{w}-N_{ke}+1)}{N},
	\label{CNN_rmps}
\end{align}
where $N_f$ is the number of filter, $N_{ke}$ is the kernel size for four input channels corresponding to in-phase and quadrature symbols for X and Y 
polarizations. In case information from multiple subcarriers are fed as input 
to the ANN core $N_f$ should be scaled accordingly. The convolutional layer is 
assumed to have zero padding with single stride and dilation.
For \LSTM~network, consider the input sequence length for each direction as 
$N_s = k + N$ where $k = t - (N_{ke}-1)/2$ is the extra symbol length at each 
side of \LSTM~input. In this case, the combined RMs for forward and backward 
\LSTM~is given by:
\begin{align}
	LSTM_{RM} = \frac{2N_sN_h(4(N_f+N_h)+3)}{N},
	\label{LSTM_rmps}
\end{align}
where $N_h$ is the hidden size.
Finally, for an MLP with single hidden layer at the output of \LSTM~network, the RM is described by:
\begin{align}
	MLP_{RM} = n_m.2N_h + 2n_m.
	\label{MLP_rmps}
\end{align}
where $n_m$ is the hidden layer size. In case MLP contains more than one hidden layer, extra multiplications should be added, accordingly. Furthermore, in the absence of any hidden layer, $MLP_{RM} =4N_h$ where 4 is 
the multiplication of 2 directions of \LSTM~and 2 outputs in I and Q for each 
output symbol. 

Note that to obtain complexity for each structure in Section~\ref{NN_struct_sec}, we need to calculate and accumulate the RMs associated to ANN cores in the equalization path for all subcarriers.
Thus, we mainly use the number of real multiplications per super-symbol (RMpS) 
as the complexity metric for each realization of an architecture. Note that 
super-symbol denotes the combined output symbols for all digital subcarriers across one polarization at each time instance. While we limit the scope of this paper to a \DMB~system with four sub-carriers, this metric enables us to further compare the results with other single carrier and \DMB~transmission systems that operate on the similar baudrate and tailored for the same throughput in future studies.

\section{ANN Structures for NLC in \DMB}\label{NN_struct_sec} 
\subsection{\DMBCombinedFC \ (\DMBCombinedFCabbr)}
\begin{figure}[t]
	\centering
	\begin{subfigure}{.45\textwidth}
	\centering
	\includegraphics[width=1\textwidth]{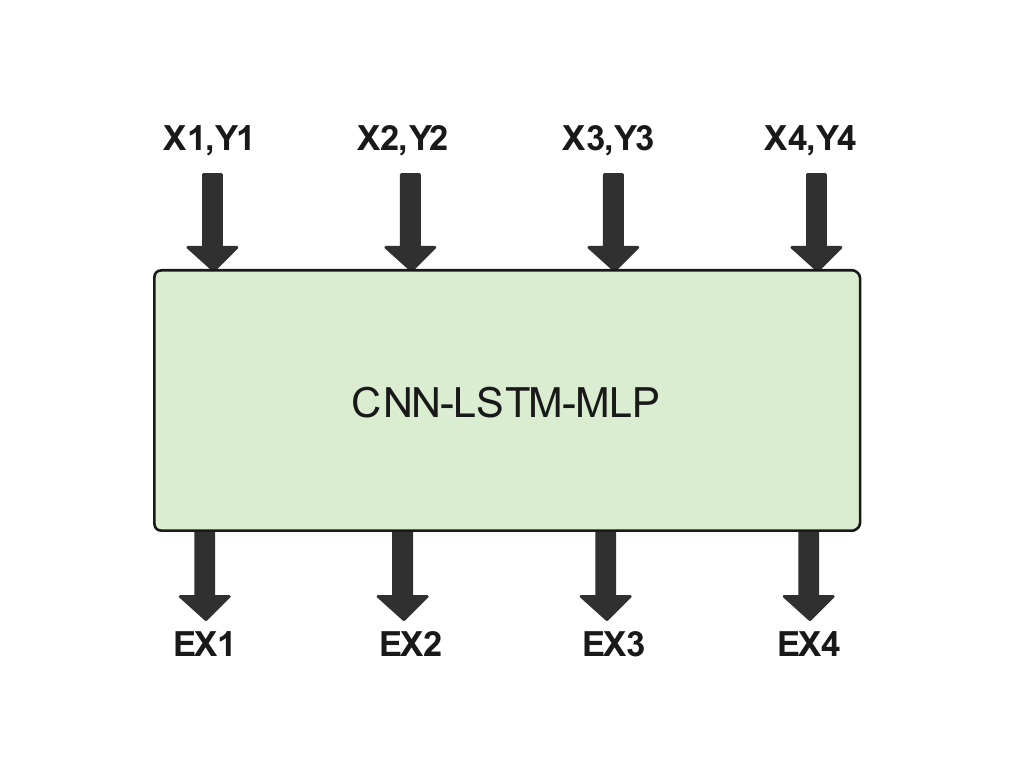}
	\caption{\DMBCombinedFC}
	\label{diag:combined}
	\end{subfigure}
	\hfill
	\begin{subfigure}{.45\textwidth}
	\centering
	\includegraphics[width=1\linewidth]{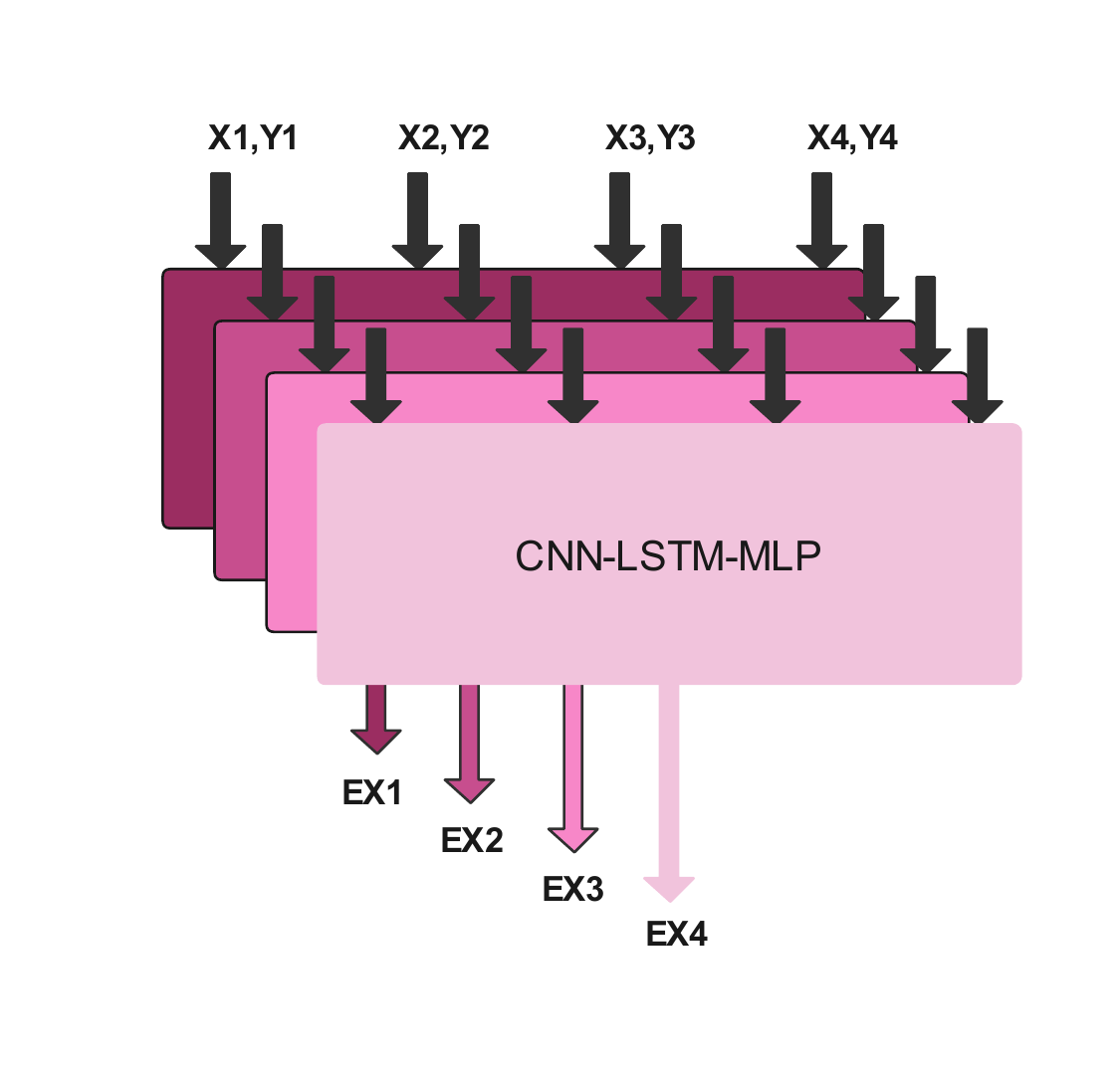}
	\caption{\DMBCombinedPerTone}
	\label{diag:combinedPerTone}
	\end{subfigure}
	\caption{ANN-NLC structures: (a) \DMBCombinedFC, and (b) \DMBCombinedPerTone}
\end{figure}

First structure for joint NLC in \DMB~is a fully-connected black-box approach 
that contains only one ANN core. This single ANN core is tasked to provide 
nonlinear distortion estimates for all subcarriers of one polarization using a 
window of received symbols from all subcarriers in both polarizations (as 
depicted in Fig.~\ref{diag:combined}). 
Note that employing \DMBCombinedFCabbr~model which lacks any enforced 
structure that separates the iSPM and iXPM nonlinear contributions could be seen as a double-edge sword. 
In one hand, this increases the number of training parameters compared to a specialized physics-informed ANN where a predetermined structure is enforced on the ANN architecture. 
On the other hand, by lack of adding any structure on the construct of the network, we may allow maximum entanglement of iSPM and iXPM features through different layers of the ANN core. 
This can potentially lead to higher efficiency by allowing the network to avoid duplicating terms that could be shared in the absence of a single and fully-connected structure. However, there is always the possibility that the ANN core structure may not be inherently powerful enough for the underlying nonlinear mechanism to learn all the appropriate features even by allowing higher complexity realizations. This could severely limit the performance, especially in the absence of adequate training data and defeat this purpose.

\subsection{\DMBCombinedPerTone ~(\DMBCombinedPerToneabbr)}
In order to obtain a subcarrier-based structure and parallelize the model, we allocate a separate ANN core to estimate the nonlinear distortion for each subcarrier output. Note that similar to \DMBCombinedFCabbr, ANN cores in \DMBCombinedPerToneabbr~still operate on input information from all subcarriers. This design is illustrated in Fig.~\ref{diag:combinedPerTone}. 
The motivation here is to employ separate and smaller cores per subcarrier in order to be more effective in fine-tuning the model parameters. This is 
important since inner and outer subcarriers may experience different balances 
of iSPM and iXPM nonlinear distortions. Also, in terms of flexibility, in case there are inactive subcarriers due to the network throughput demands, such as hitless capacity upgrades or in P2MP scenarios, the parallel design in \DMBCombinedPerToneabbr~could be more efficiently deployed compared to the single connected core architecture of \DMBCombinedFCabbr. 
However, one potential drawback for this structure is that utility sharing between equalization paths of different subcarriers is prevented.

\subsection{\DMBvTwo~(\DMBvTwoabbr)}
\begin{figure}[t]
	\centering	
	\begin{subfigure}{.55\textwidth}
		\includegraphics[trim={1cm 0cm 1cm 1cm},clip, width=.95\textwidth]
		{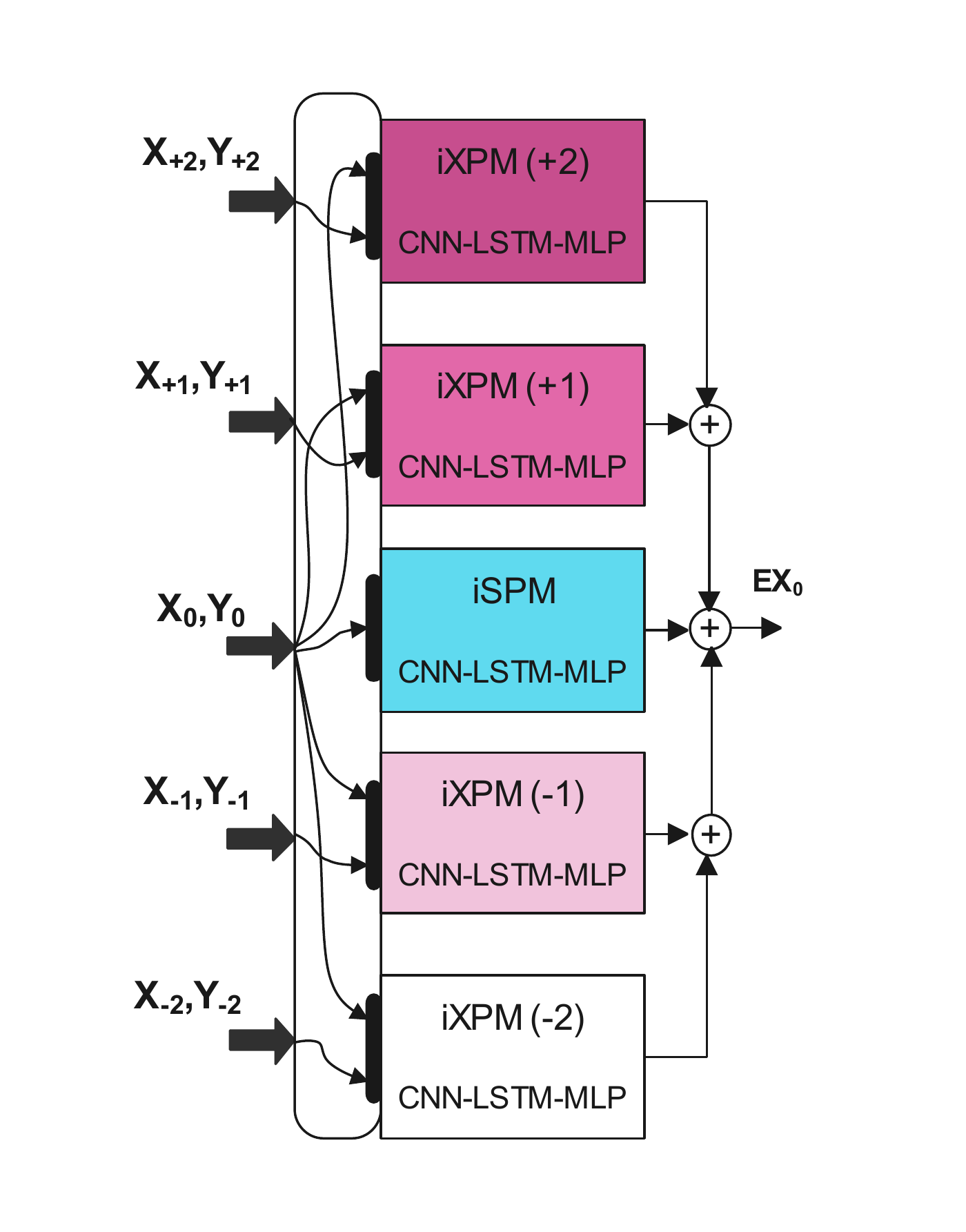}
		\caption{\DMBvTwo~realization for each subcarrier.}
		\label{diag:DMBv2_1}
	\end{subfigure}
	\begin{subfigure}{.4\textwidth}
		\centering		
		\includegraphics[trim={2cm 1.5cm 1.5cm 1cm},clip, 
		width=1\textwidth]
		{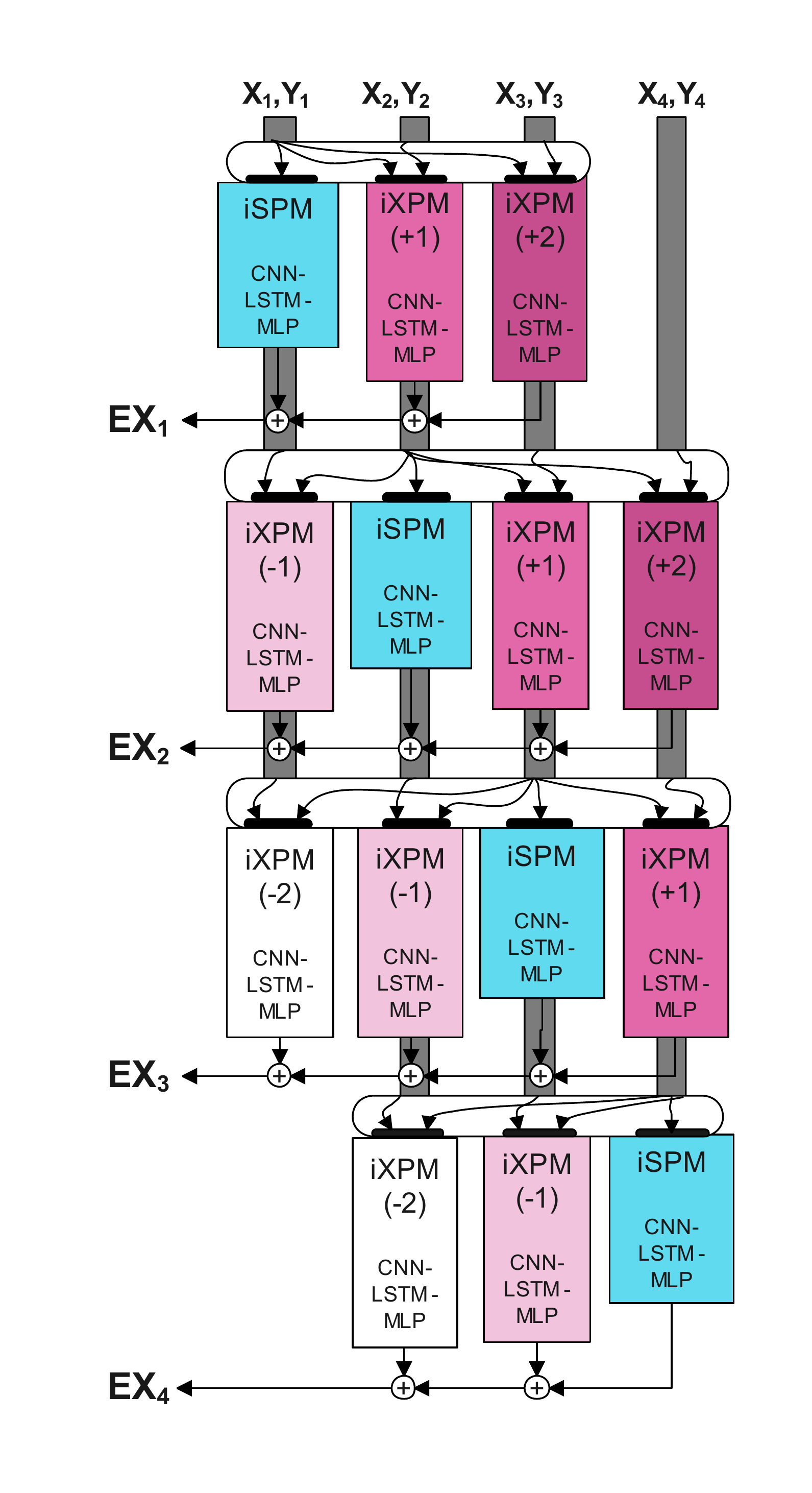}
		\caption{\DMBvTwo~design for \DMB~ANN-NLC.}
		\label{diag:DMBv2_2}
	\end{subfigure}
	\caption{Structural design of ANN-NLC using  \DMBvTwo. (a) illustrates the trained cores and (b) illustrates the implementation for a 4-subcarrier system.}
	
\end{figure}

We move on from the black-box approach to design more efficient and flexible ANN-NLC models by using the insights from perturbation analysis of fiber nonlinear propagation.
Specifically, the underlying mathematics behind iSPM triplet coefficients in perturbation analysis weakly relies on the absolute position of a subcarrier in the spectrum \cite{pert1}
\footnote{This statement is technically accurate in the absence of higher order linear distortion terms such as dispersion slope, but could be sufficiently accurate even in presence of such terms in practical systems.}.
Also, iXPM nonlinearity mechanism relies on the relative position of target and interfering subcarriers. Hence, only a small set of 
iSPM and iXPM cores can be trained and multiple instances of these trained cores get deployed as needed in the equalization path.
Furthermore, since iXPM contributions are more pronounced among neighboring 
subcarriers, smaller and more efficient networks can be deployed by involving 
only iXPM contributions of immediate neighboring subcarriers.
Fig.~\ref{diag:DMBv2_1} illustrates a set of ANN cores that can be trained for \DMBvTwoabbr~design where one iSPM and four iXPM cores are trained to model the intra- and inter-subcarrier nonlinearities for up to two neighboring subcarriers from each side.
Note that the input to an iSPM core is a window of the target subcarrier symbols while the iXPM cores employ symbols from both target and interfering 
subcarriers. 

Let us look at an implementation of \DMBvTwoabbr~model that considers iXPM 
nonlinearities of up to two neighboring interfering subcarriers from 
each side of every output subcarrier. The block diagram for this modular NL 
equalizer is depicted in Fig.~\ref{diag:DMBv2_2} where four iSPM cores 
compensate self nonlinearities originated from each subcarrier. Moreover, two 
inner and outer subcarrier pairs additionally employ three and two iXPM cores, 
respectively. 
Note that ANN cores with similar color share same layouts and weights leading to more efficient training specifically with limited data. 
Provided that channel parameters and subcarrier bandwidth and spacing remain the same, additional cores with learned weights and biases from this example can be deployed for systems with higher number of subcarriers. 
With a proper training strategy, the proposed structure allows us to separate 
iSPM and iXPM contributions and informatively direct computational 
resources to the best route. This is evident in the numerical results where we explore moving beyond iSPM compensation for various modular designs.

Another advantage of this modular design can be seen in certain scenarios, such as hitless capacity upgrades or in P2MP scenarios, wherein certain subcarriers could be turned-off. In this case, \DMBCombinedPerToneabbr~and specially \DMBCombinedFCabbr~models trained with all subcarriers may not be efficiently utilized as the statistics of inputs to ANN core(s) for to the deactivated subcarrier(s) would be vastly different from the training. 
Additionally, it would be almost impossible to effectively identify and disable routes within ANN that correspond to the absent subcarriers to save power or reduce penalty.  However, a modular design can be readily reconfigured to accommodate such scenarios by deactivating the equalization paths corresponding to absent subcarriers, leading to a flexible and power-efficient deployment.

\subsection{\DMBModular \ (\DMBModularabbr)}
The next step in the evolution of ANN-NLC for \DMB~is rooted in two observations.
First, the perturbation analysis \cite{pnlc3,pnlc4,pnlcDMB} suggests that the 
iXPM perturbation coefficients $C_{m,n}^{(-\ell)}$ governing the interaction of subcarrier $i$ and its $\ell$'s neighbor on right $i+\ell$ are similar to those of subcarrier $i$ and its $\ell$'s neighbor on left  $i-\ell$, provided that we employ a 
simple transformation, i.e.,
\begin{equation}
C_{m,n}^{(-\ell)} =  {C_{-m,n}^{(\ell)}}^*.
\end{equation}
where $m$ and $n$ are the symbol indices. 
Additionally, since these perturbation coefficients mainly rely on the relative position of subcarriers, the similarity can be extended to subcarrier 
$i+\ell$ and its $\ell$'s neighbor on the left, $i$. Note that the iXPM$(+\ell)$ and 
iXPM$(-\ell)$ cores in \DMBvTwoabbr~for $i$ and $i+\ell$ subcarriers, respectively, are solely fed 
by inputs from these two subcarriers.
This hints to potential computational savings by merging iXPM$(+\ell)$ and iXPM$(-\ell)$ cores in \DMBvTwoabbr~that operate on same subcarriers into a super core iXPM$(\pm \ell)$ and potentially obtain a more efficient structure that preserve similar performance levels with a lower complexity.
%

The output features of these super-cores along with the appropriate iSPM 
features are passed to separate MLP modules prior to aggregation for each subcarrier. 
Note that MLP layers are detached from the ANN cores in this design and a set of $2\ell+1$ MLP modules are trained in this approach to model 
integration of iSPM features and up to $2\ell$ iXPM core features involving 
neighboring subcarriers. The trained MLP modules are appropriately instantiated in the inference 
path for each subcarrier.
\begin{figure}[b]
	\centering
	\includegraphics[trim={1cm 1.5cm 1cm 1.25cm},clip, width=.8\textwidth]
	{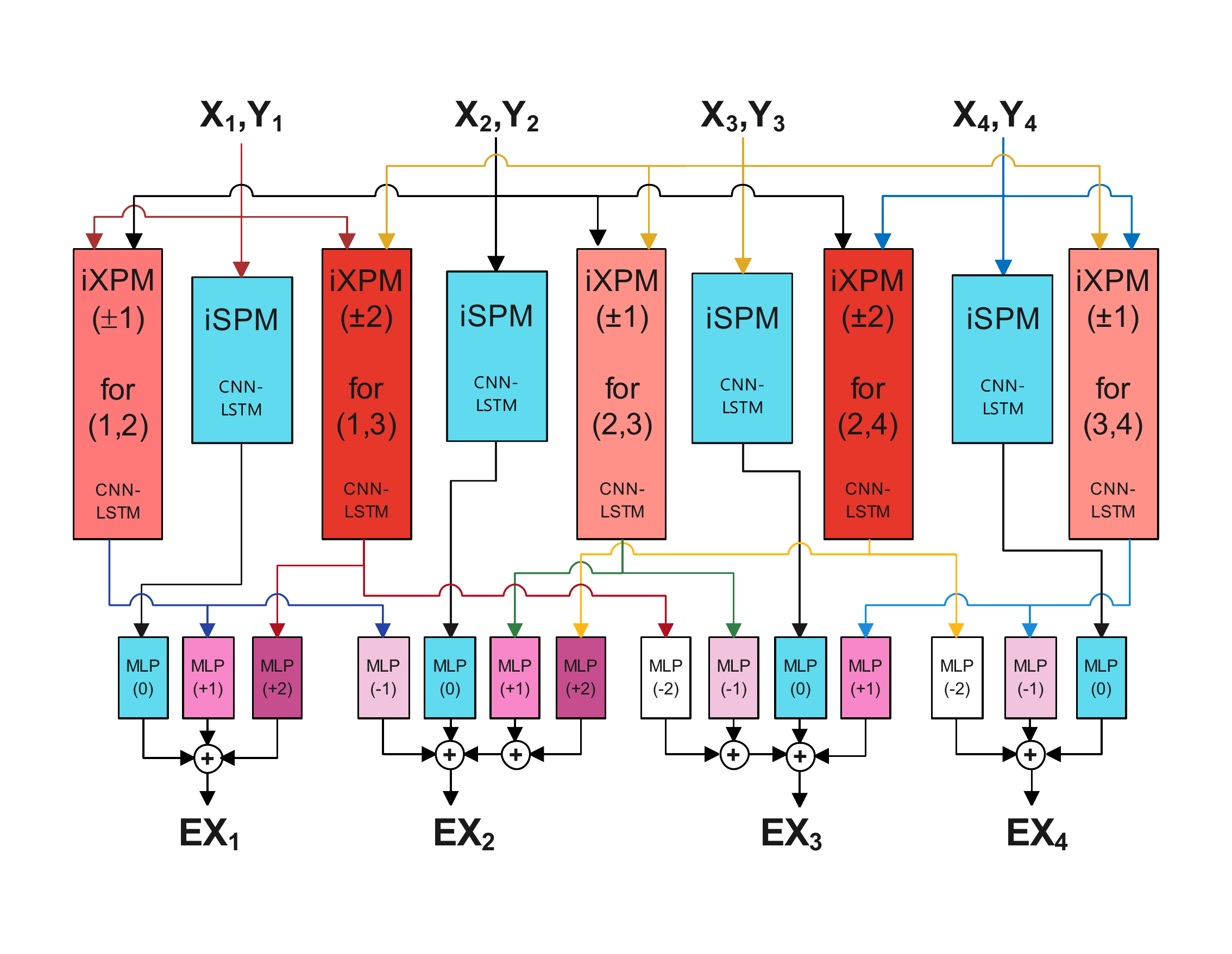}
	\caption{\DMBModular~design for \DMB~ANN-NLC.}
	\label{diag:DMBModular}
\end{figure}

In summary, potential performance versus complexity tradeoff advantages of  
\DMBModularabbr~could be of two fold. 
First, merging cores that are believed to contain significant amount of shared 
computations for feature generation can increase model efficiency. 
Second, reducing the distinct parameters of a network by replicating trained modules can greatly improve training efficiency and result in more generalized models. 
Also as mentioned before, the modular design provides additional 
flexibility in crafting more intelligent solutions for different network operational scenarios.  
Fig.~\ref{diag:DMBModular} shows a block diagram for this model with four subcarriers and $\ell=2$.

\section{Numerical Results}\label{sys_dmb}
\subsection{System Model}
\begin{figure}[t]
	\centering	
	\includegraphics[width=1\textwidth]{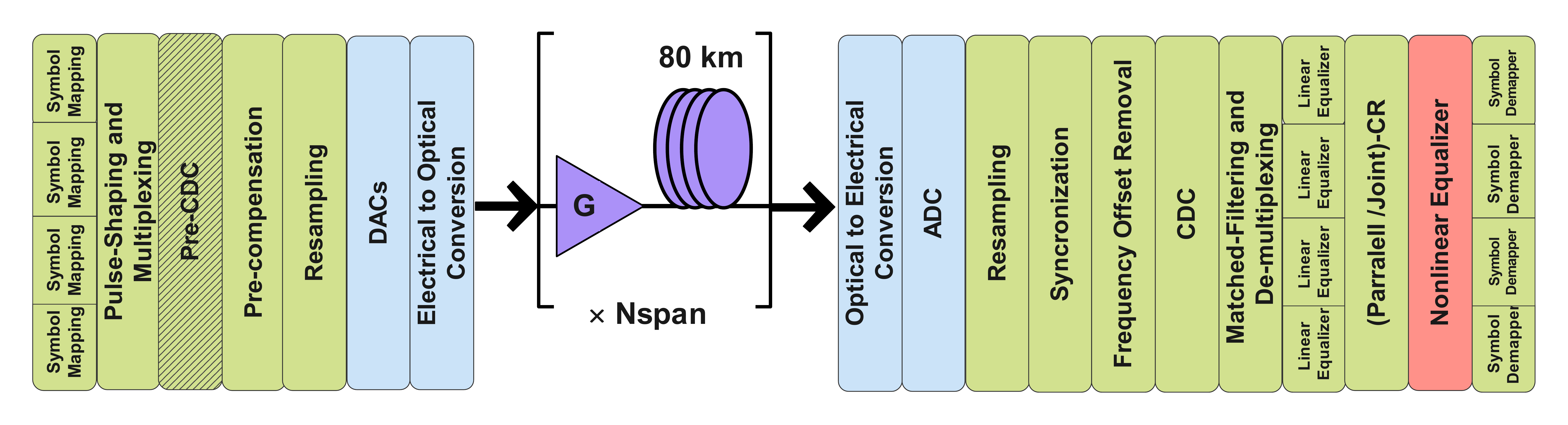}
	\caption{System model for \DMB~system.}
	\label{diag:DMB_system_2210}
\end{figure}

The simulation setup includes typical Tx, channel, and Rx modules for a \DMB~transmission scenario. 
To focus on fiber nonlinearity, we consider ideal electrical
components and Mach-Zhender modulator. Also, DACs/ADCs are ideal with no quantization or clipping effects.
The dual polarization fiber channel is modeled by split-step Fourier method \cite{SSFM} with adaptive step-size and maximum nonlinear phase-rotation of 0.05 degree to ensure sufficient accuracy.
At the Rx side, the sequence output from carrier recovery (CR) are used to train and evaluate the nonlinear equalizer. 
Standard DSP algorithms are employed for detection and processing of the received signal at the Rx. 
The block diagram for such system is depicted in Fig.~\ref{diag:DMB_system_2210}.
Note that, to keep the ability of conventional coherent receiver for phase correction under correlated phase-noise (which is coming from nonlinear propagation in our case), we deployed the carrier recovery before ANN-NLC module.
This ensures that the linear equalization provides nonlinear phase compensation 
capability of a coherent receiver without a dedicated NLC equalizer. Hence, the neural network compensation gain is given on top of the best linear performance. 

To evaluate and optimize different algorithms, we focus on a single-channel 
\DMB~system operating at 32 Gbaud with four subcarriers and uniform 16QAM 
modulation format. 
The signal at each subcarrier is digitally generated using root-raised cosine pulse shape with a roll-off factor of $1/16$. 
The link consists of 40 spans of standard single-mode fiber of 80km length, 
followed by optical amplifiers with $NF=6$ dB noise figure. 
Furthermore, for the most of numerical results we consider a symmetric dispersion map, in which 50\% of total dispersion is digitally pre-compensated at the transmitter side.
This in turn allows us to simplify the diagrams and avoid unnecessary 
complications at this stage. 
Section~\ref{Rx-side CDC} is devoted to extension of this design to other dispersion maps where we provide ANN-NLC structures optimized for a post-CDC scenario.
The training and evaluation of models are performed using at $2$ dBm launch power. 
This is close to the optimal launch power when DBP at 2 Sa/sym and 1 and
2 steps-per-span are employed to benchmark these results. 
Note that for this setup, a Q-factor of $Q=7.88$ dB can be obtained at the optimal launch power of $1$ dBm in the absence of a fiber-nonlinearity compensation.


\subsection{ANN Optimization Workflow}
All the models here are trained and evaluated based on the simulation data 
using $2^{18}$ symbols per digital subcarrier. The training and evaluation data are 
generated from pseudo-random streams and different generator seeds using  
permuted congruential generator (PCG64).
Also, $20 \%$ of the training dataset was set aside for validation of the model 
during the training process. 
Root minimum squared error (RMSE) between the model outputs and the difference 
between transmitted symbols and the received values constitutes the loss, which 
is used in the back-propagation process to update the model coefficients. 
All models were trained using Adam optimizer with learning-rate $=0.001$ for at 
least $200$ epochs, unless terminated by the early-stopping mechanism that 
tracks the validation loss and prevents over-fitting.
We mainly used mini-batches of length $512$ in obtaining these results. Minor 
performance differences were observed by exploring mini-batch sizes as low as 
$128$, and as high as $2048$ provided that learning-rate and number of 
epochs were optimized accordingly. 
Additionally, we employed a learning-rate scheduler that reduces the 
learning-rate by $20\%$ when loss stops reducing for 10 epochs. 
For each model, best coefficients associated to the least validation loss 
across all training epochs were saved at the end of the training stage.

\begin{table}[t]
	\small	
	\renewcommand{\arraystretch}{1}
	\caption{List of hyper-parameters for ANN core that operates on a sequence length $T=2t+1$ with $t \in [5:40]$. 
		\label{table:params}} 
	\centering 
	\begin{tabular}{c l c c} 
		\hline\hline 
		Layer & Learn-able Parameters & value / sweep range  \\ [0.5ex] %
		\hline 
		CNN & $num\_layers$ & 1  \\ 
		& $num\_output\_channels$ & [10:200] \\
		& $kernel\_size$ & [5:30] \\
		\\
		LSTM & $num\_hidden\_state$ & [10:300] \\
		& $num\_output\_features$ & [10:300] \\
		\\
		MLP & $num\_hidden\_layers$ & [0:2] \\
		& $layer\_size$ & [10:100] \\
		\hline 
	\end{tabular}
\end{table}

In order to explore performance versus complexity tradeoff, more than a thousand models for each design are trained and tested in this work for different block-sizes.
Table \ref{table:params} illustrates the list of the ANN-Core hyper-parameters 
and their sweeping ranges. The sweeping resolution of each parameter within 
each participating ANN core are individually adjusted for each model structure. 
We use the scatter plots reflecting the performance-complexity of different realizations of each model based on a common test dataset obtained from a separate transmission simulation using noise and bit sequences of different random-number generator algorithms and seeds. 
The envelope associated to be best performing models at various 
complexity constraints are generated in order to compare different 
architectures.

\subsection{Numerical Results Comparison}
\begin{figure}[t]
	\centering
	\includegraphics[trim={1cm 1cm 1cm 1cm},clip, width=.9\textwidth]
	{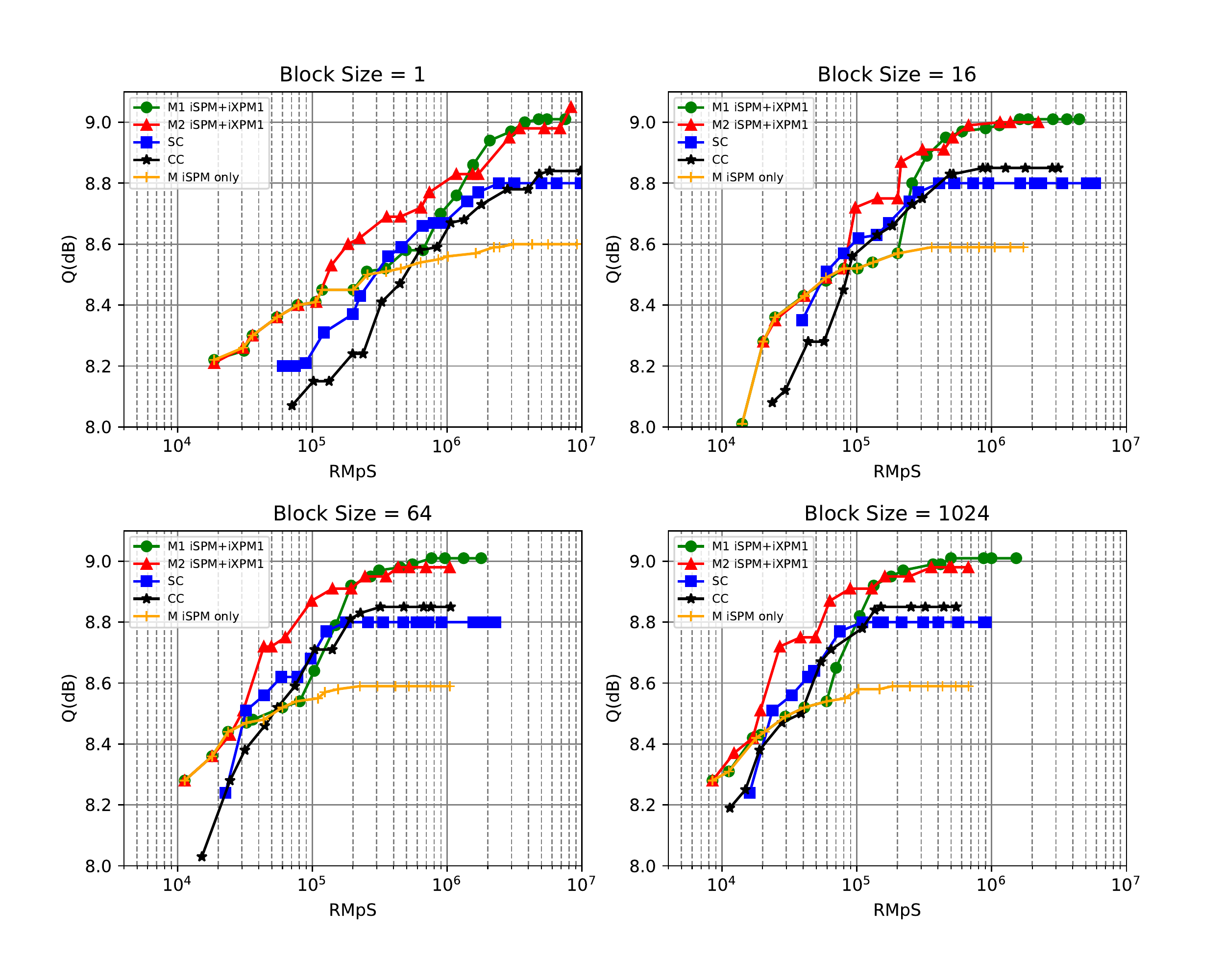}	
	\caption{A comparison of performance as a function of RMpS  amongst 
		different explored ANN-NLC solutions for \DMB.}
	\label{fig:DMB_Compare}
\end{figure}

In this part, we provide performance versus complexity tradeoff comparison of various optimized ANN equalizers for different block-sizes. 
Fig.~\ref{fig:DMB_Compare} illustrates the inference cost of various models in terms of RMpS. 
From ANN design point of view, it is important to efficiently allocate additional complexity in order to improve performance since majority of 
the models demonstrate subpar efficiency. As an example, increasing hidden size of LSTM may not be an efficient strategy to improve the performance if filter-tap size $k$ is not large enough to capture the nonlinear memory.

It can be seen that using a separate ANN core per each subcarrier did not significantly change the outcome of \DMBCombinedPerToneabbr~compared to \DMBCombinedFCabbr. Their best performance remains around 8.8 dB. The performance-complexity tradeoffs for these models remain very similar for different block sizes.
One can clearly observe various advantages of modular solutions compared to 
the black-box approaches represented by \DMBCombinedFCabbr~and  
\DMBCombinedPerToneabbr.  Both 
modular solutions offer clear superiority in both low and high complexity 
regions while \DMBModularabbr~structure, specifically,  demonstrates superior 
performance complexity trade-off across all complexity regions among all 
structures. Note that the performance for iSPM compensation is capped around 8.6 dB. 
Employing additional cores to compensate for intra-subcarrier nonlinearities 
due to immediate neighboring  subcarriers $\ell=1$ from each side (iXPM1) can significantly increase the 
maximum 
performance to around 9 dB, unlocking 0.4 dB gain compared to the iSPM 
compensation at 2 dBm launch power.
We further explored another scenario by incorporating iXPM contributions of two subcarriers from each side. 
However, the results are omitted as we did not observe a meaningful additional performance gain for this scenario.
This result can be corroborated by findings in Section \ref{Rx-side CDC}, where we demonstrate the perturbation coefficients corresponding to the iXPM contributions of the second neighbors for this setup showing that the magnitude of these coefficients are around 10dB lower than iXPM contributions from the immediate neighbors.

Note that the best performance obtained from modular solutions are generally 0.2 dB higher than  \DMBCombinedFCabbr~and  \DMBCombinedPerToneabbr. This suggests that these solutions can more efficiently learn from a limited training data due to a more generalized  structure with fewer trainable parameters. 
The trade-off between performance and complexity in mid-tier performance 
regions with Q around 8.6 dB is particularly noteworthy where non-modular 
designs can compete with \DMBvTwoabbr. Note that this region is the onset of switching away from iSPM only NLC to incorporate iXPM nonlinearities from immediate neighboring subcarriers. 
This suggests that \DMBCombinedFCabbr~and \DMBCombinedPerToneabbr~architectures 
can converge to a moderately efficient structures by internally sharing 
resources of iXPM compensation between neighboring subcarriers. This type of 
resource sharing is one of the main distinctive features of \DMBModularabbr~model compared to \DMBvTwoabbr~that reflects in its superior efficiency in this region.

\begin{figure}[]
	\centering
	\includegraphics[trim={1cm .5cm 1cm 1cm},clip, width=.7\textwidth]
	{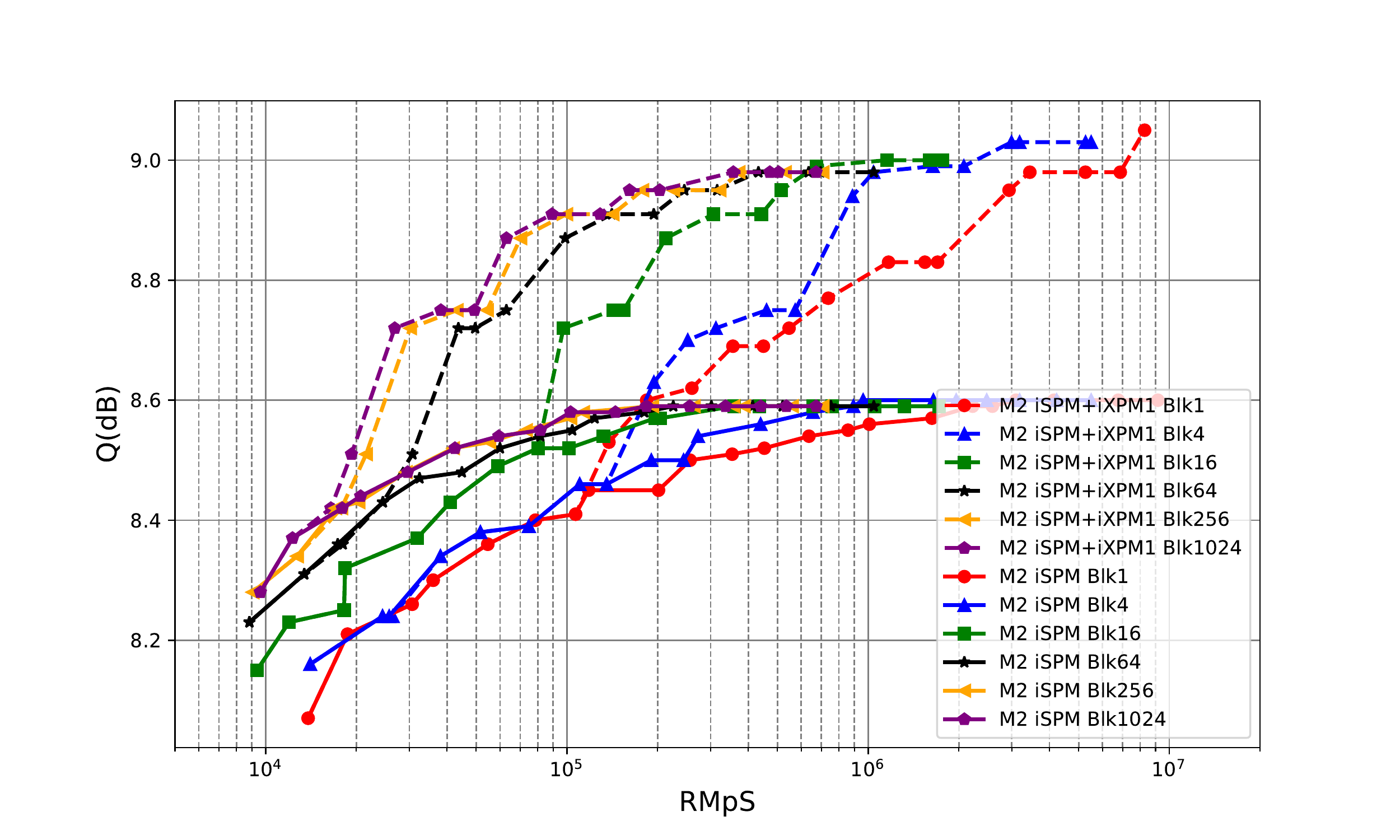}
	\caption{Impact of block-size on performance vs. complexity of the best \DMBModularabbr~models. }
	\label{fig:DMB_Modular_BLOCK}
\end{figure}

In order to demonstrate advantages of block-processing, performance versus complexity evaluations for different block-sizes are illustrated in Fig.~\ref{fig:DMB_Modular_BLOCK} for \DMBModularabbr~model as an example. 
A substantial complexity reduction for a very minimal performance loss can be obtained by parallelization of the trained ANN core and deploy the 
solution with a block-size $N>1$ provided that the model is sufficiently generalized in the training stage. 
In high-performance region ($Q>8.8$), we can achieve a complexity 
reduction by a factor of $20$ for $N=1024$. 
However, the complexity advantages shrinks in lower performance regions 
(ex. factor of $5$  for $Q\sim8.4$) where best models generally have lower filter-tap size and incorporate less nonlinear memory.

\begin{figure}[b]
	\centering
	\includegraphics[width=.55\textwidth]
	{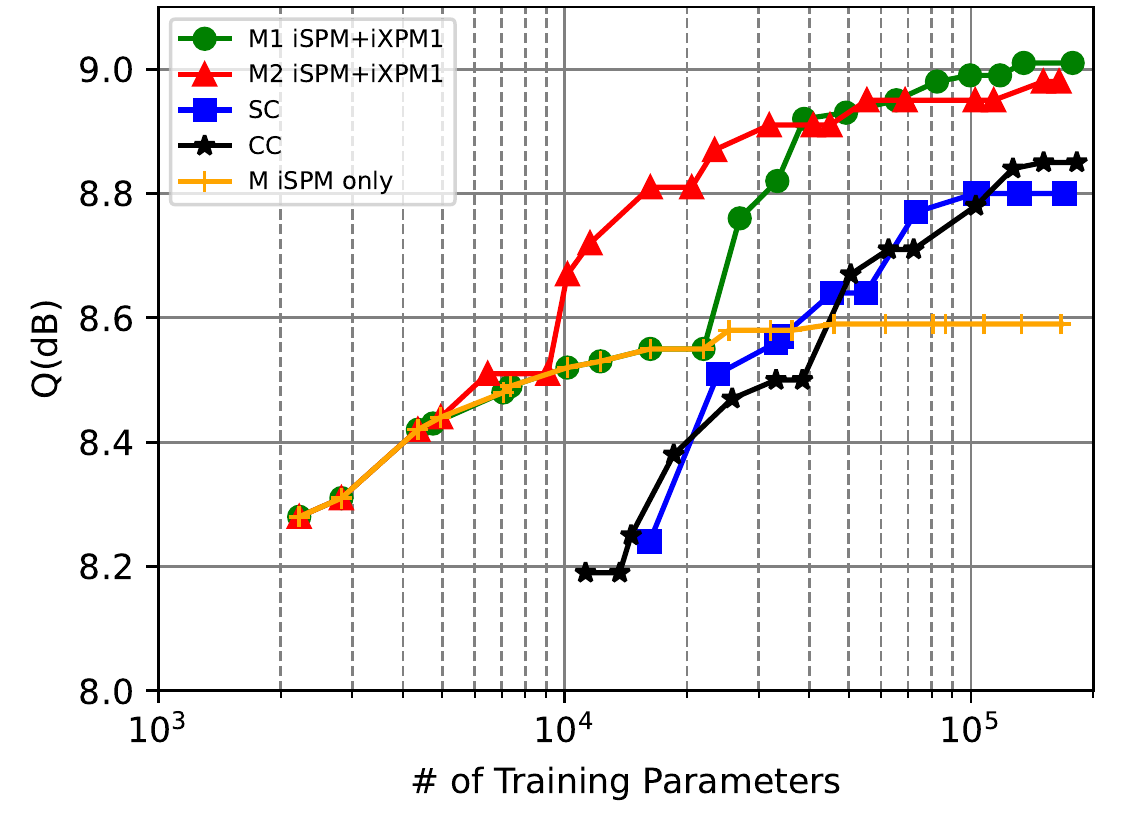}
	\caption{A comparison of performance as a function of number of training 
		parameters amongst different explored ANN-NLC solutions for \DMB.}
	\label{fig:DMB_Compare_training}
\end{figure}

Next, performance envelopes for all models as a function of the number of training parameters are depicted in Fig.~\ref{fig:DMB_Compare_training}. The number of training parameters is related to the memory requirements to store and retrieve model parameters as link configuration is modified over-time. This metric can also measure the efficiency of a model to provide a certain performance level with the least independent parameters which is also closely tied with generalization of the ANN. 
%
For a mid-tier performance of around 8.6 dB, the modular solutions generally 
demonstrate approximately 2 to 4 times lower number of parameters compared to \DMBCombinedFCabbr~and  \DMBCombinedPerToneabbr.
Note that \DMBCombinedFCabbr~and  \DMBCombinedPerToneabbr~solutions supposedly have access to all subcarrier information and are not limited to iSPM+iXPM1 architectures of  \DMBvTwoabbr~and \DMBModularabbr. However, this assumed \emph{advantage} results in a significant loss for \DMBCombinedFCabbr~and \DMBCombinedPerToneabbr~solutions if the number of training parameters are below 40,000. 
We attempted to close this performance gap by increasing the number of epoches for non-modular solutions and further optimizing the learning rates without much success. This may indicate that practical ANN design in presence of various limitations and constraints for this problem is far from a plug and play approach and requires careful design using insights from the physical model.

\begin{figure}[t]
	\centering
	\includegraphics[trim={1cm .5cm 1cm .5cm},clip, width=1.03\textwidth]
	{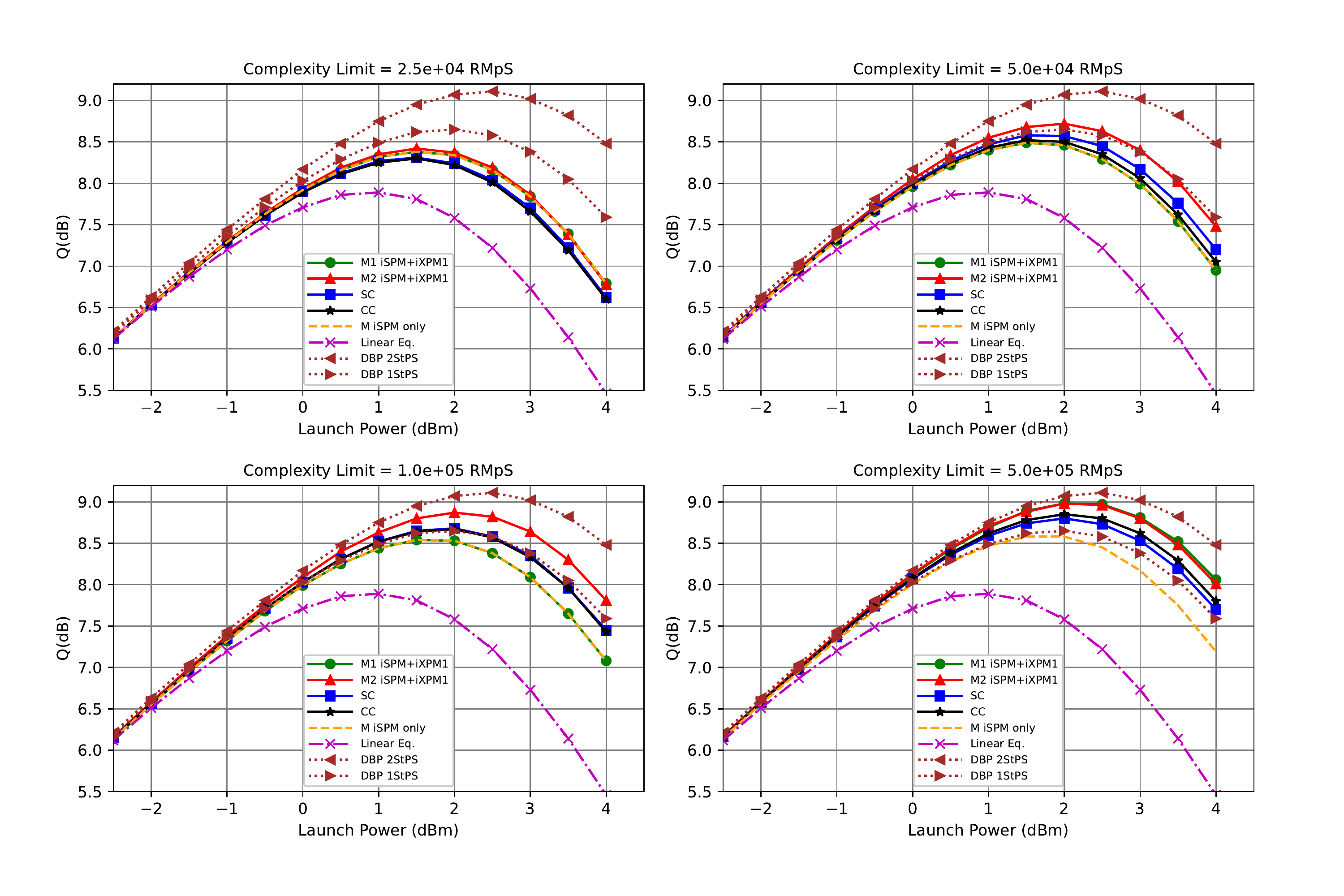}
	\caption{Performance of different ANN-NLC solutions as a function of optical launch power for given complexity constraint budgets.}
	\label{fig:DMB_Power_sweep}
\end{figure}

Finally, we explore the applicability of proposed models on similar links with different optical launch powers.
Fig.~\ref{fig:DMB_Power_sweep} illustrates the performance as a function of 
optical launch power where ultiple graphs are presented for best models obtained with different complexity budget constraints. As stated earlier, all models where trained 
at $2$ dBm optical launch power. Note that selected models from all structures  
demonstrate good generalization and can provide nonlinear performance gain on a wide range of launch powers, spanning from linear regime to deep nonlinearity. We provide 
DBP performance plots with different number of steps per span (StPS) to 
benchmark proposed ANN-NLC structures. Note that, the complexity comparison with 
other NLC methods such as DBP is not performed here since a fair comparison 
requires development of efficient hardware-friendly versions of ANNs after model compression, pruning, and weight quantization which is beyond the scope of this paper.

\section{Impact of Dispersion Map} \label{Rx-side CDC}
\begin{figure}[t]
	\centering
		\includegraphics[trim={.5cm .5cm .5cm 1cm},clip, width=.95\textwidth]
	{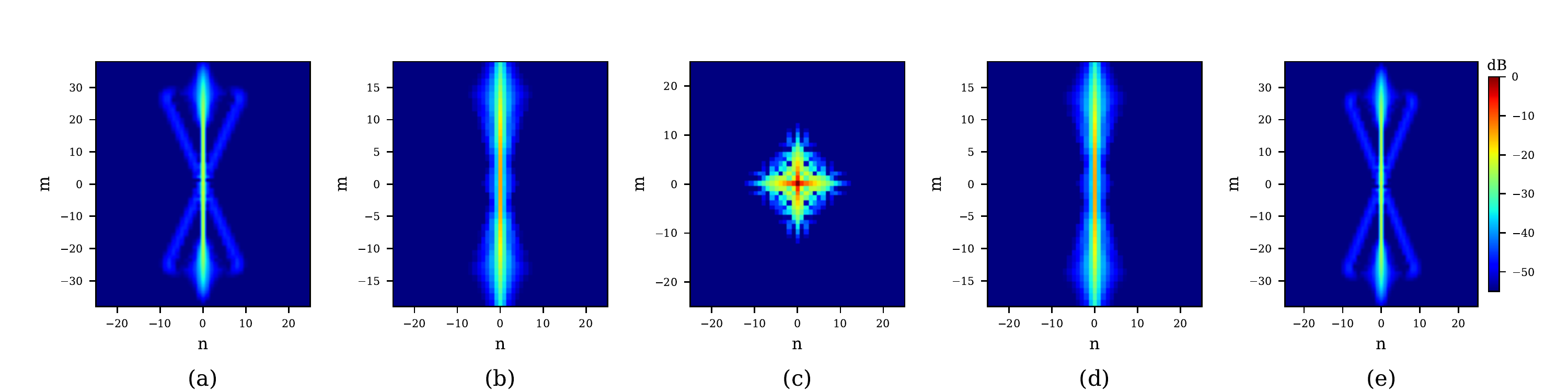}
	\caption{Magnitude of iSPM ($\ell=0$) and iXPM ($\ell\not=0$)  perturbation coefficients 
		$C_{m,n}^{(\ell)}$ for the \DMB~simulation setup with sym-CDC: (a) $\ell=-2$, 
		(b) $\ell=-1$, (c) $\ell=0$, (d) $\ell=1$, (e) $\ell=2$. \label{diag:pert_symCDC}}
	\centering
	\includegraphics[trim={.5cm .5cm .5cm 0cm},clip, width=.95\textwidth]
	{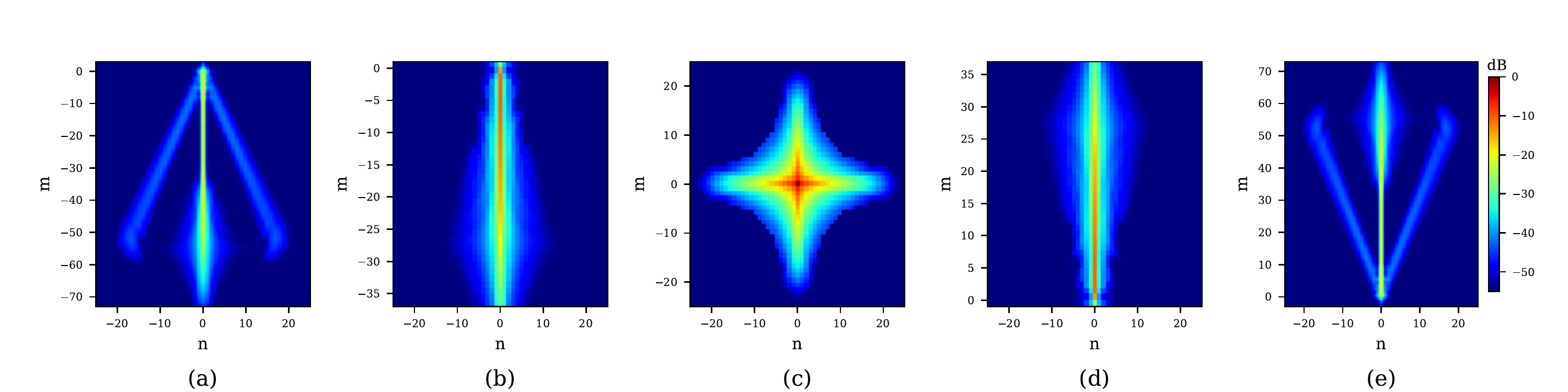} 
	\caption{Magnitude of iSPM ($\ell=0$) and iXPM ($\ell\not=0$)  perturbation coefficients 
		$C_{m,n}^{(\ell)}$ for the \DMB~simulation setup with post-CDC: (a) $\ell=-2$, (b) 
		$\ell=-1$, (c) $\ell=0$, (d) $\ell=1$, (e) $\ell=2$. \label{diag:pert_posCDC}}
\end{figure}

\begin{figure}[b]
	\centering
	\includegraphics[trim={1.6cm 1.6cm 1.5cm 1.6cm},clip, width=.35\textwidth]
	{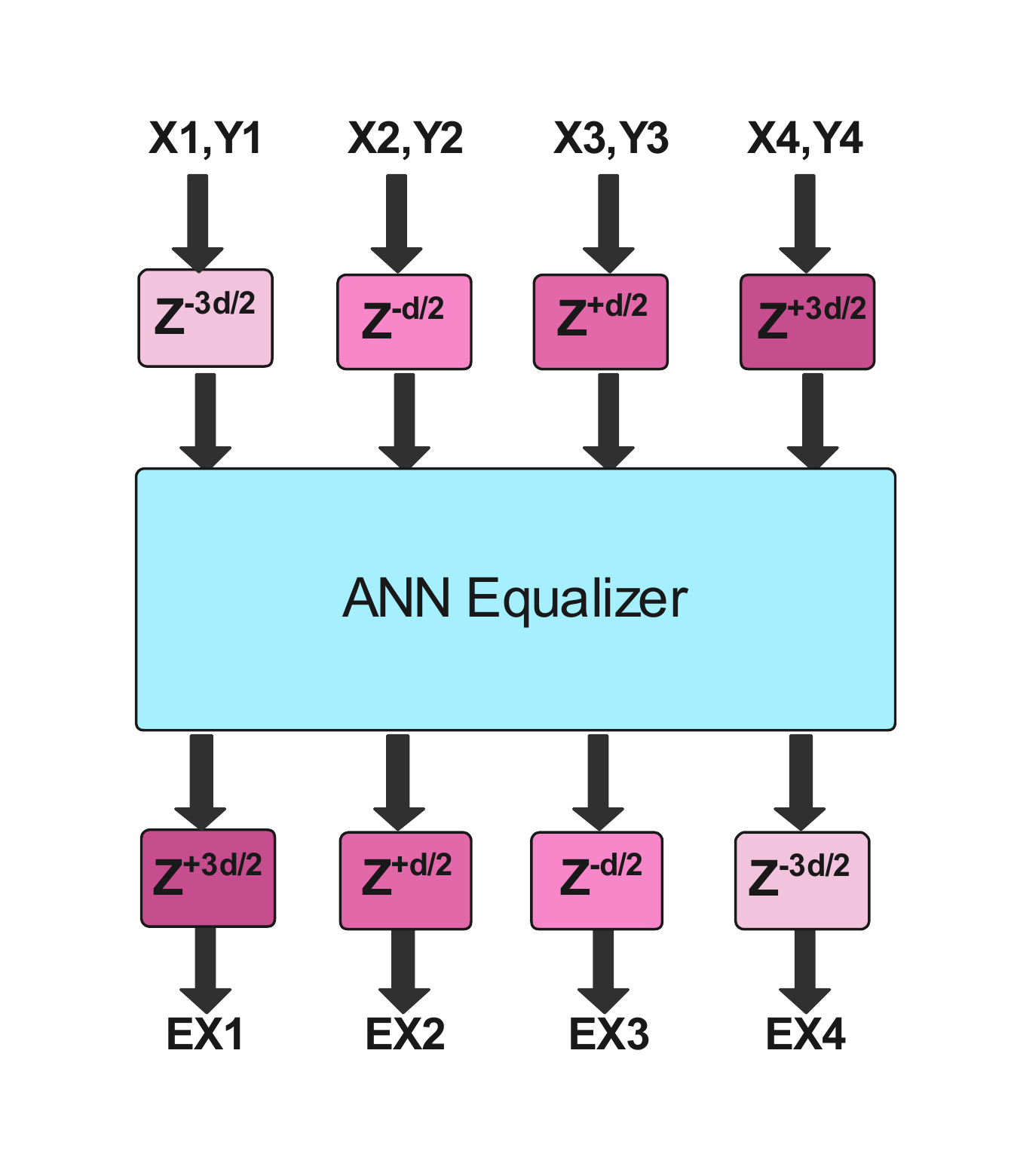}
	\caption{Delay adjustment for post-CDC dispersion map.}
	\label{Diag:DMB_postCDC}
\end{figure}

So far, we have shown application of ANN-NLC equalizers in transmission 
scenarios with symmetric dispersion map. 
As depicted in Fig.~\ref{diag:pert_symCDC}, the windows for symbols of interest from target 
and interfering subcarriers for iSPM and iXPM triplet features for a symmetric 
dispersion map are symmetric  
around reference symbols. This is the main reason that symmetric windows of 
soft values are selected as input to iSPM an iXPM cores in the previous 
designs. However, in presence of an asymmetric dispersion map, such as post 
dispersion compensation, the regions for iXPM features of most significance are neither symmetric nor centered around the reference symbol from interfering 
subcarrier as shown in Fig.~\ref{diag:pert_posCDC}. 
Hence, one needs to adjust the input features for each iXPM core according to the dispersion-induced group-delay between the involved subcarriers. 
Another approach is to introduce delay lines in the input and output of the ANN equalizer and maintain a symmetric input windows for the ANN cores.
Specifically, to ensure proper operation of the equalizer in this case, we 
introduce a progressive delay amounting to half of the dispersion-induced group delay between subcarriers prior to the ANN equalization. To reverse this impact, 
another delay-line is added at the out of the ANN equalizer. Note that the window size for each iXPM core needs to be as large as the maximum group-delay between 
associated subcarriers. This ensures symbols that impacted the target symbol 
are appropriately involved.
Fig.~\ref{Diag:DMB_postCDC} illustrates a block diagram for this solution. 

\begin{figure}[htbp]
	\centering
	\includegraphics[trim={.1cm .2cm .4cm .2cm},clip, width=.9\textwidth]
	{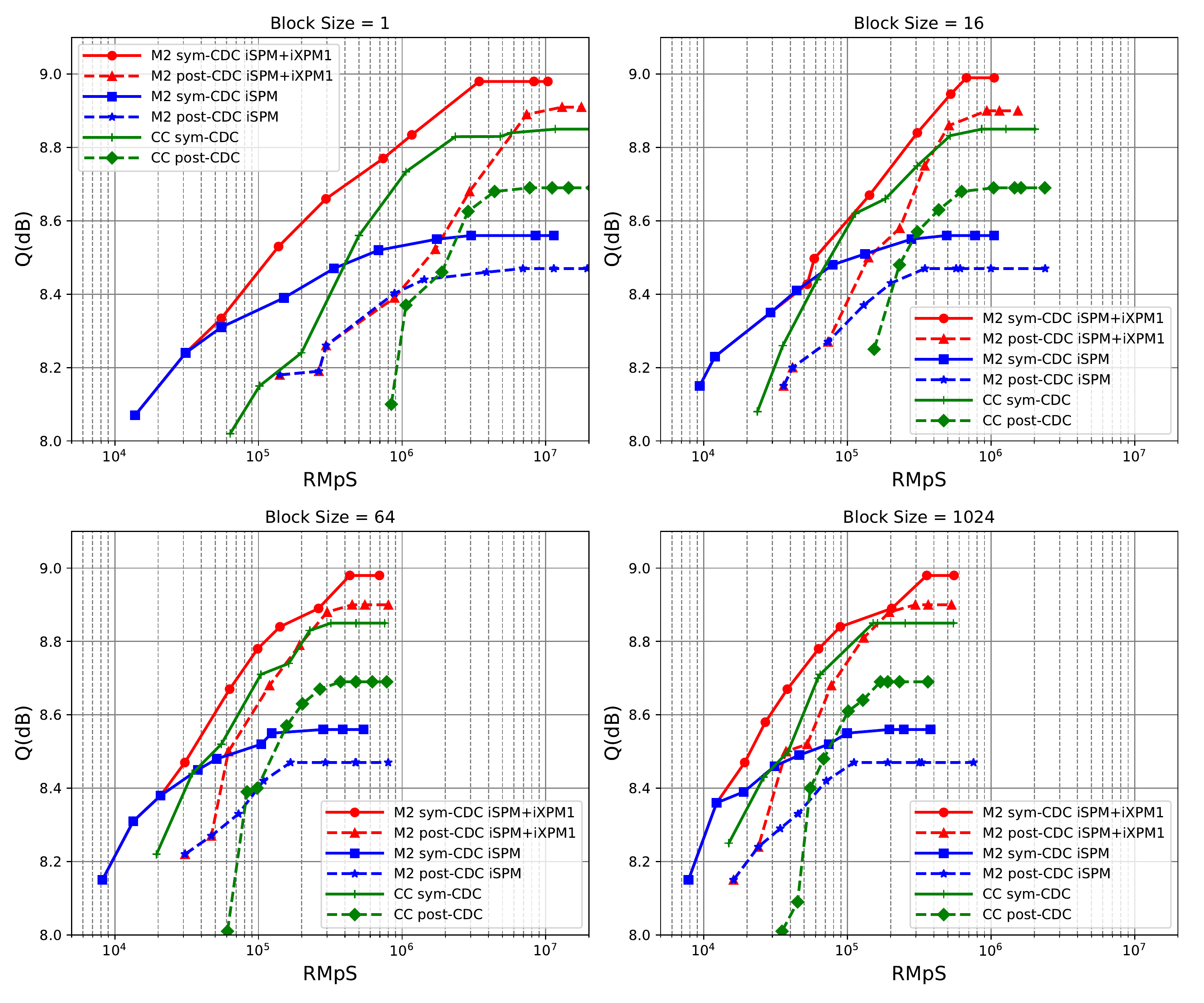}
	\caption{Comparison on the impact of dispersion map on effectiveness of ANN-NLC using  envelope associated to best performing models at different block-sizes.}
	\label{fig:DMB_CDC}
\end{figure}

We have modified the simulation setup to provide a performance
comparison of selective ANN equalizers between symmetric- and post-CDC
in Fig.~\ref{fig:DMB_CDC}.
Similar trends are observed for \DMBCombinedFCabbr~and 
\DMBModularabbr~solutions with post-CDC, showing the applicability and 
effectiveness of the proposed solution. 
Note that similar performance gains are achieved by switching from iSPM to 
iSPM+iXPM1 nonlinear equalization for these schemes.
Additionally, we observe that the complexity of all NLC solutions with post-CDC 
is higher than their respective counterpart with symmetric CDC for a given performance 
level. This can be attributed to the larger memory of iSPM and iXPM 
nonlinearities in the link with post-CDC. This is corroborated by comparing the domain and magnitude of perturbation coefficients presented in Fig.~\ref{diag:pert_symCDC} and Fig.~\ref{diag:pert_posCDC}.

\section{Conclusion}\label{conclusion_sec} 
In this work, we studied different ANN approaches in compensation of 
intra-channel nonlinearities in \DMB~systems.
By training and evaluating various models over a comprehensive grid of parameters, we explored performance versus complexity tradeoff of each approach and discussed their scalability, potentials and weaknesses.
Starting from back-box approaches in designing ANN models, we gradually moved towards modular designs inspired by perturbation analysis of fiber nonlinearity. This approach proved to be more efficient in training and producing better models with a given training data, 
as well as inference complexity and model storage requirements. We further 
demonstrate a pragmatic approach to adapt the proposed solutions to links with 
asymmetric dispersion maps. While these networks were exclusively designed for 
fiber nonlineairty compensation, a similar approach can be further studied in 
the context of component nonlinearity compensation in \DMB~systems.

Note that all these designs can be further optimized by looking at other avenues. 
Notable approaches such as weight pruning, quantization and also future extension to quantization-aware training in form of quantized and binary neural networks can be explored to drastically reduce complexity of these models. Despite this, we believe that our presented study provides a fair comparison and good starting step towards that path by focusing on the macro design of the ANN equalizers tailored to the characteristics of the fiber nonlinearity distortion mechanism in multi-subcarrier systems.

\bibliographystyle{IEEEtran}
\bibliography{OpExpress_full}

 \end{document}